\documentclass[rnote]{aa} % for the letters 

\usepackage{lscape}
\usepackage{graphicx}
\usepackage{txfonts}
\usepackage{natbib}

\usepackage{color}

\begin{document}

   \title{An infrared diagnostic for magnetism in hot stars\thanks{Based on observations obtained at the Southern Astrophysical Research (SOAR) telescope, which is a joint project of the Minist\'{e}rio da Ci\^{e}ncia, Tecnologia, e Inova\c{c}\~{a}o (MCTI) da Rep\'{u}blica Federativa do Brasil, the U.S. National Optical Astronomy Observatory (NOAO), the University of North Carolina at Chapel Hill (UNC), and Michigan State University (MSU). }}

   \author{M.~E.~Oksala\inst{1}
          \and
          J.~H.~Grunhut\inst{2}
                  \and
          M.~Kraus\inst{3,4}
          \and
          M.~Borges~Fernandes\inst{5}
         \and
          C.~Neiner\inst{1}
          \and
          C.~A.~H.~Condori\inst{5}
          \and
          J.~C.~N.~Campagnolo\inst{5}
          \and
         T.~B.~Souza\inst{5}
          }

   \institute{LESIA, Observatoire de Paris, CNRS UMR 8109, UPMC, Universit\'{e} Paris Diderot, 5 place Jules Janssen, 92190 Meudon, France\\
              \email{mary.oksala@obspm.fr}
         \and
             European Southern Observatory, Karl-Schwarzschild-Str. 2, D-85748 Garching, Germany
             \and
             Astronomick\'{y} \'{u}stav, Akademie v\v{e}d \v{C}esk\'{e} republiky, Fri\v{c}ova 298, 251 65 Ond\v{r}ejov, Czech Republic
                          \and
             Tartu Observatory, T\~oravere, 61602 Tartumaa, Estonia
                        \and
             Observat\'{o}rio Nacional, Rua General Jos\'{e} Cristino, 77 S\~{a}o Cristov\~{a}o, 20921-400, Rio de Janeiro, Brazil
                         }

   \date{Received ; accepted }

 \abstract{Magnetospheric observational proxies are used for indirect detection of magnetic fields in hot stars in the X-ray, UV, optical, and radio wavelength ranges. 
 To determine the viability of infrared (IR) hydrogen recombination lines as a magnetic diagnostic for these stars, we have obtained low-resolution (R$\sim$1200), 
near-IR spectra of the known magnetic B2V stars HR 5907 and HR 7355, taken with the Ohio State Infrared Imager/Spectrometer (OSIRIS) 
attached to the 4.1m Southern Astrophysical Research (SOAR) Telescope.
Both stars show definite variable emission features in IR hydrogen lines of the Brackett series, with similar properties as those found in optical spectra, including the derived location of the detected magnetospheric plasma.
These features also have the added advantage of a lowered contribution of stellar flux at these wavelengths, making circumstellar material more easily detectable.  
IR diagnostics will be useful for the future study of magnetic hot stars, to detect and analyze lower-density environments, and to detect magnetic candidates in areas obscured from UV and optical observations, increasing the number of known magnetic stars to determine basic formation properties and investigate the origin of their magnetic fields.
}


   \maketitle
%
%________________________________________________________________

\section{Introduction}

\begin{figure}
\centering
\includegraphics[width=65mm]{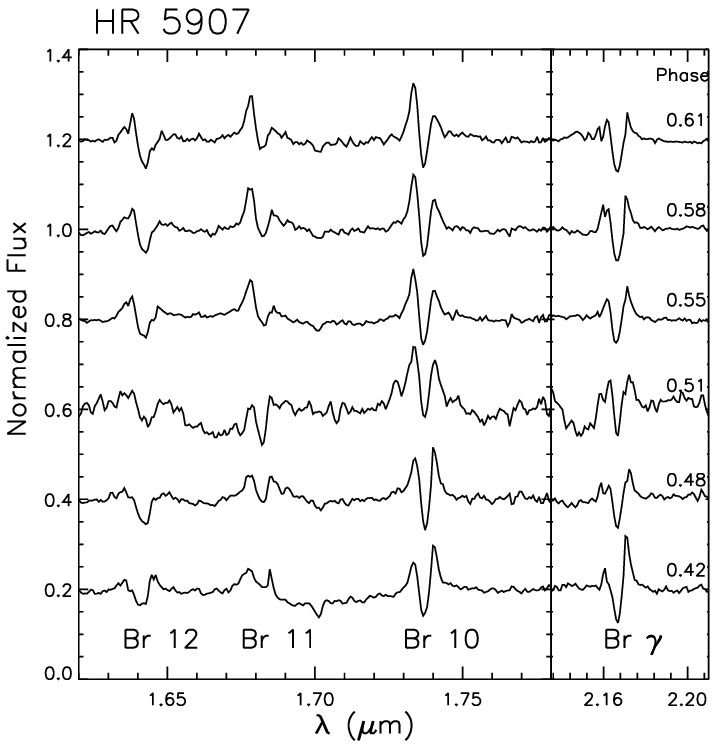}
\includegraphics[width=65mm]{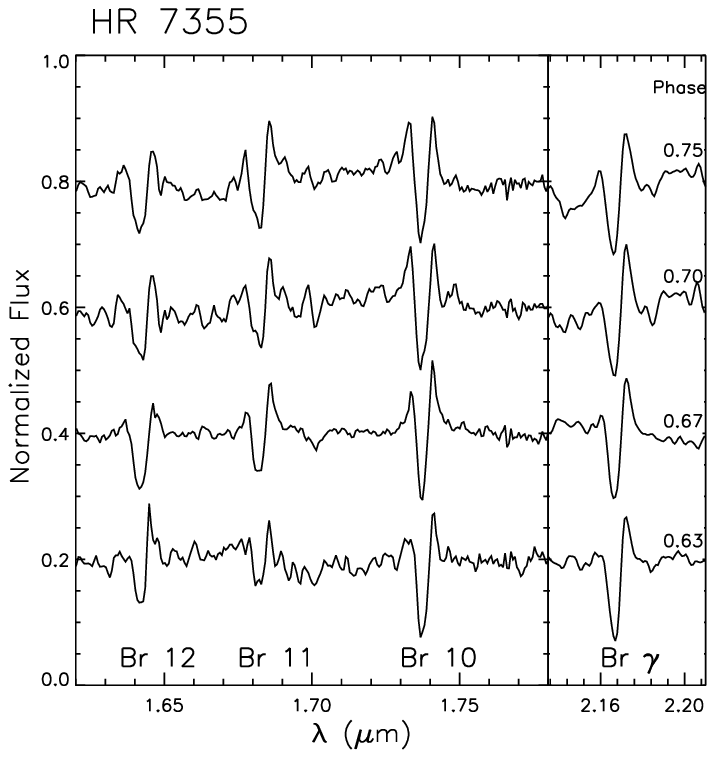}
\caption{Time-series of low-resolution (R$\sim$1200), near-IR spectra of the He-strong B2V stars HR 5907 (\textit{top}) and HR 7355 (\textit{bottom}).  
Stellar rotational phases for each spectrum are indicated on the right, and the hydrogen recombination lines are identified along the bottom.
}
\label{spec}
\end{figure}

Extensive work over the past four decades has revealed the presence and general properties of large-scale magnetic fields in a subset of hot stars \citep[see, e.g.,][]{Landstreet:1978aa, Borra:1979aa, Donati:2002aa}.   Long-term studies demonstrate that these magnetic structures remain stable over long periods \citep[see, e.g.,][]{Oksala:2012ab,Silvester:2014aa}, making them ideal laboratories to examine the interaction of such fields with the moderate to strong stellar winds of OB stars.  
Competition between the strengths of the field and wind in the circumstellar environments of magnetic hot stars creates a magnetosphere, 
or a region above the stellar surface occupied by ionized gas \citep{Babel:1997ab,ud-Doula:2002aa}.  As the stellar wind drives material from the surface, 
the plasma follows the field lines from opposite footpoints to the top of the magnetic loop, where the two streams collide, cool, and then either are held up by centrifugal forces or fall back onto the surface of the star.

This material can be detected in a number of ways and thus can be used as a proxy for direct magnetic detection 
through spectropolarimetric observations \citep[see, e.g.,][and references therein]{Petit:2013aa}.  
{In the optical, the cooler and denser post-shock material trapped in the stellar magnetosphere is detected in the hydrogen recombination lines, 
typically H${\alpha}$ lines, although the emission pattern depends on the magnetic and stellar properties \citep[i.e., field strength, mass-loss rate, 
stellar geometry;][]{Oksala:2012ab,Rivinius:2013aa,Grunhut:2012ab}.  Other physical manifestations of the magnetosphere can be detected in observations of X-ray \citep[e.g.,][]{Gagne:2005aa},
UV \citep[e.g.,][]{Henrichs:1993aa}, and radio \citep[e.g.,][]{Trigilio:2004aa}, typically either presenting as peculiar features or variability.  However, to date, there have been no dedicated studies of infrared (IR) spectral features 
to obtain analogous diagnostics for hot stars.}

%Luminous, hard, variable X-ray emission traces shocks created by the channeled, magnetically-confined wind, such as in the 
%cases of $\tau$ Sco \citep{Cassinelli:1994aa} and $\theta^{1}$ Ori C \citep{Gagne:2005aa}.  
%However, there is currently no empirical model for this diagnostic, and there are several counter examples where the X-ray spectrum is soft and/or without variation \citep[e.g., HD 191612;][]{Naze:2010aa}.  
%Optical photometry can also experience variability as a result of either circumstellar material 
%eclipsing the surface of the star \citep[e.g.,][]{Walborn:1976aa} or inhomogeneous surface chemical abundances \citep[e.g.,][]{Krticka:2007aa}.
 
%In the UV wavelength range, ionized gas is detected via resonance lines, such as Si~{\sc iv}, C~{\sc iv}, and N~{\sc v}.  In the presence of a 
%magnetically confined stellar wind, these spectral lines show rotational modulation \citep{Henrichs:1993aa,Kaper:1994aa}, and in some cases super-ionization 
%\citep{Petit:2011aa,Oskinova:2011aa}.  These variations are currently used as a diagnostic to search for
%magnetic hot star candidates \citep{Henrichs:2012aa}.  
%Variable radio emission is observed in many higher mass magnetic stars, as a result of the interaction between the stellar wind and magnetic field outside of the Alfv\'en radius 
%\citep[see e.g.,][]{Trigilio:2004aa,Leto:2006aa,Leto:2012aa}.  

{IR spectral studies of normal OB stars \citep[e.g., ][]{Hanson:2005aa,Najarro:2011aa} 
show hydrogen recombination lines, either in emission or in absorption.  
%These and other studies aim to uncover the spectral features of these stars, model them using NLTE codes such as CMFGEN and FASTWIND, and determine how to use these spectra to determine spectral type and study wind properties, such as mass-loss and clumping. 
%In these cases, emission is typically seen as a consequence of an extended atmosphere (i.e. for supergiants) or a dense stellar wind. 
{Analyses of Be stars found emission combined with minimal photospheric contribution in IR hydrogen line profiles, 
allowing a more direct investigation of the circumstellar material than possible with optical features \citep[see,
e.g.,][]{Lenorzer:2002ab,Granada:2010aa}.  }
As magnetic OB stars also have circumstellar material and may show H${\alpha}$ emission (albeit typically weaker features), 
it would be logical to suggest that the IR spectra of magnetic stars similarly show strong emission features.
%and that these IR features could provide important insights on the magnetospheres of hot stars. 
}

With the research above as motivation, we began a survey to investigate the IR spectral properties of {known} magnetic, hot stars.  
During the course of this work, \citet{Eikenberry:2014aa} observed in their study of red giants as part of 
Sloan Digital Sky Survey III's Apache Point Observatory Galactic Evolution Experiment \citep[SDSS-III/APOGEE,][]{Eisenstein:2011aa}
two early B-type stars in the H band as telluric standards, {but found that they contained strong hydrogen line emission with peak velocities 
far outside what would normally be expected for Be stars,
%.  Upon further investigation, the authors found that, when compared with an H-band spectrum of the prototypical magnetic 
%Bp star $\sigma$ Ori E, the three objects appeared quite similar
leading the authors to conclude that these two stars were candidate magnetic B stars. 
The findings of \citet{Eikenberry:2014aa} provided justification to continue development of our work. }

{Here, we present our pioneering pilot study into the viability of IR hydrogen recombination lines as a magnetic diagnostic for massive stars, 
examining time-series of two well-known magnetic chemically peculiar stars, HR 5907 and HR 7355. }

%Section 2 describes the observations obtained and their reduction process.  
%We present the results of our study in Section 3, and discuss the possibilities for the use of this type of diagnostic.  Concluding remarks are offered in Section 4.

\begin{table}
\caption{Observation summary}             
\label{observe}      
\centering          
\begin{tabular}{lcc} 
\hline\hline 
Object  &   HJD & Phase\\
\hline 
HR 5907    &   2456462.60178            &  0.42\\
                 &   2456462.63069              &  0.48\\
                 &   2456462.64967              &  0.51\\
                 &   2456462.66552              &  0.55\\
                 &   2456462.68113              &  0.58\\
                 &   2456462.69877              &  0.61\\
\hline
HR 7355    &  2456462.72116             &  0.63\\
                 &   2456462.73978              &  0.67\\
                 &   2456462.75889              &  0.70\\
                 &   2456462.78514              &  0.75\\
\hline\hline 
\end{tabular}
\end{table}

\section{Targets and observations} 

For this preliminary investigation, we chose two targets, the He-strong B2V stars HR 7355 and HR 5907, both known to possess strong magnetic fields and circumstellar material.  
A comprehensive study of HR 5907 by \citet{Grunhut:2012ab} revealed a strong, 
primarily dipole magnetic field with a strength of 10.4 kG and extremely rapid rotation ($\varv \sin i =  290$ km s$^{-1}$, P$_{\rm{rot}} = 0.508276$\,d).  
The authors also found strong, variable H$\alpha$ emission and a variable, optical photometric light curve, both indicating the presence of 
plasma trapped in a co-rotating (centrifugal) magnetosphere.  Similarly, HR 7355 had been suggested to be a magnetic star by \citet{Rivinius:2008aa}, 
based on the appearance of H$\alpha$ emission in its spectrum and a variable, optical photometric light curve.
This notion was confirmed by \citet{Oksala:2010aa}, and \citet{Rivinius:2010aa,Rivinius:2013aa}, 
which indicated that HR 7355 indeed possesses a primarily dipolar magnetic field with a strength of 11.6 kG.  The star was also found to be rapidly rotating with P$_{\rm{rot}} = 0.52144$\,d
and $\varv \sin i =  310$ km s$^{-1}$.  Ultimately, the strength of the optical magnetospheric signatures of the two stars makes them ideal targets for identifying corresponding IR diagnostics.

\begin{figure}
\centering
\includegraphics[width=40mm]{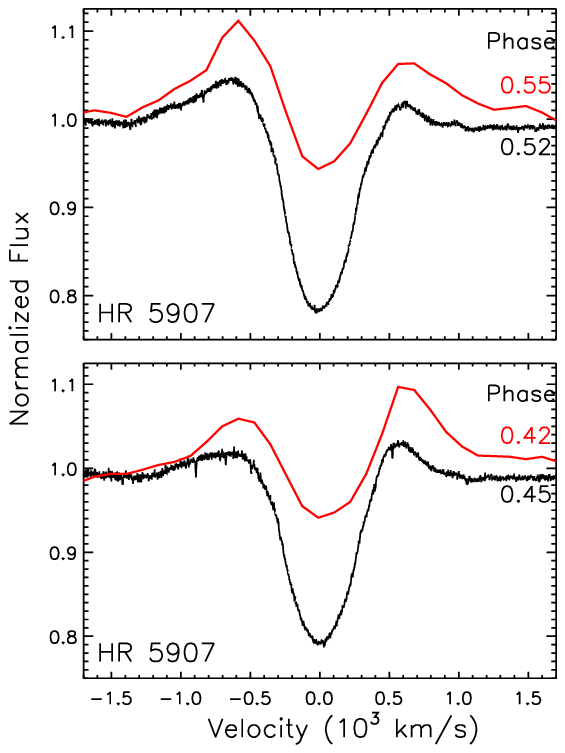}
\includegraphics[width=40mm]{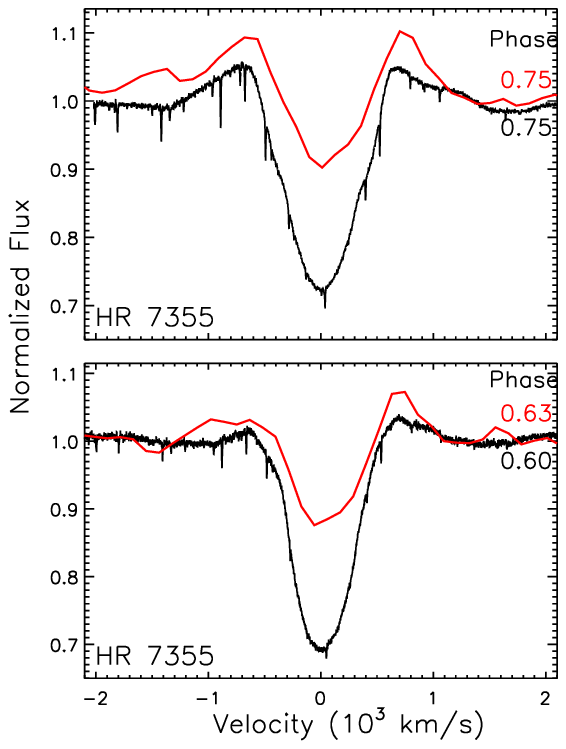}
\caption{Comparison of optical (H$\alpha$, \textit{black}) and IR (Br10, \textit{red}) spectral features.  {Left:} HR 5907 at two different rotational phases, $\sim0.45$ and $\sim0.55$ (phases noted to the right).   These two phases occur just before and just after the H$\alpha$ emission maximum \citep{Grunhut:2012ab}. 
{Right:} HR 7355 at phases $\sim0.60$ and $\sim0.75$.  A peak in H$\alpha$ emission occurs at phase 0.78 \citep{Rivinius:2013aa}.
%In both cases, the variability of the red and blue emission peaks between the two phases are closely matched.
%Note also, the decreased stellar flux contribution in the IR features.
}
\label{speccomp}
\end{figure}

To demonstrate the feasibility of such studies, we have obtained low-resolution (R$\sim$1200), near-IR spectra, taken with the Ohio State Infrared Imager/Spectrometer (OSIRIS) 
attached to the 4.1 m Southern Astrophysical Research (SOAR) Telescope.  The data were obtained during the night of 2013 June 18
with a 1'' slit width and in cross-dispersed mode, to simultaneously obtain $J$-, $H$-, and $K$-band spectra.  
We obtained six sets of observations of HR 5907 and four for HR 7355, with each set consisting of four exposures in an ABBA nod pattern.  Two additional spectra of HR 7355 were taken during worsening weather conditions and were in the end of too low quality to be considered.   A summary of these observations is presented in Table~\ref{observe}.  Rotational phases were computed using the ephemeris derived by \citet{Grunhut:2012ab} for HR 5907 and that derived by \citet{Rivinius:2013aa} for HR 7355.  Individual exposure times for each ABBA position were 6s for HR 5907 and 15s for HR 7355.

The data were reduced using standard IRAF
%\footnote{IRAF is distributed by the National Optical Astronomy Observatory, which is operated by the Association of Universities for Research in Astronomy (AURA) under cooperative agreement with the National Science Foundation.} 
tasks within the \textit{echelle} package, including flat-field division, sky-background subtraction, and wavelength calibration using ThAr spectra.  
A master flat image was obtained using the task xdflat within the program XDSpres, a CL script written for OSIRIS by  \citet{Ruschel-Dutra:2011aa}.
Spectra of the B9V stars HIP 78809 and HIP 93583 for HR 5097 and HR 7355, respectively, were obtained directly following the target at a similar airmass to minimize the effect of changing atmospheric conditions to create telluric templates.   
The IRAF task \textit{telluric} used these templates to remove atmospheric lines from the target spectra.   
Spectra extracted from each set of ABBA exposures were co-added to increase the final signal-to-noise ratio (S/N), which ranges from $\sim100$ to 200.

{Because of the wavelength range acquired and the location of a strong telluric region, the correction of atmospheric effects introduces uncertainties.  
The J-band spectra were unusable, and thus we proceeded using only the H and K bands.   We note that due to the low resolution of the acquired data, we are only able to identify hydrogen recombination lines, and the K band may 
still experience lingering effects, such that Br$\gamma$ may yet be contaminated.   }

\begin{figure}
\centering
\includegraphics[width=67mm]{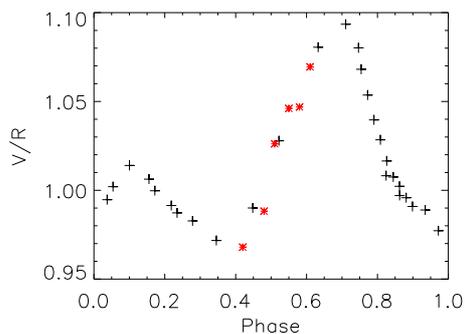}
\caption{V/R variations of observed emission in H$\alpha$ (black pluses) and Br10 (red asterisks) spectral lines for HR 5907.  }
\label{vrvar}
\end{figure}

\begin{figure}
\centering
\includegraphics[width=39mm]{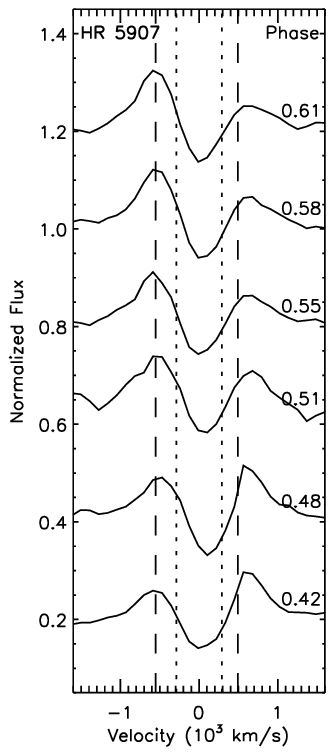}
\includegraphics[width=39mm]{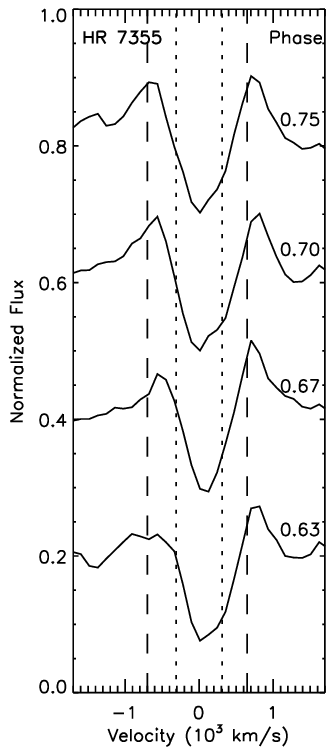}
\caption{Time-series of the Br10 line of HR 5907 (\textit{left}) and HR 7355 (\textit{right}).  
Stellar rotational phases for each spectrum are indicated to the right.  The dotted lines indicate stellar rotation velocities derived from optical spectroscopy, while 
dashed lines indicate the peak emission found for H$\alpha$ \citep{Grunhut:2012ab,Rivinius:2013aa}.
}
\label{peakvel}
\end{figure}

\section{Results and discussion}

Both HR 5907 and HR 7355 exhibit strong emission features in the line profiles of the IR hydrogen recombination lines.  
Figure~\ref{spec} shows the Br10, Br11, Br12, and Br$\gamma$ lines.  
While the data have a low resolution, the circumstellar emission is obvious at each rotational phase.  
The time series of HR 5907 (Fig.~\ref{spec}, top) only covers 20\% of the rotation period, but the spectral lines show 
obvious variability in the strengths of the red and blue emission peaks, best viewed in the Br10 line.  This variation is similar to behavior seen in H${\alpha}$ line profiles
of both of our targets.  
Although the observations of HR 7355 (Fig.~\ref{spec}, bottom) only cover 10\% of the rotational phase, the variability is easily visible.

The goal of this fundamental study is to determine whether spectral features in the IR are useful for detecting emission from a magnetosphere.  
One way to test this, given that we have precise information about the rotational modulation of the magnetic field of the star, is to compare the 
characteristics of the IR features with those of the optical features, namely H${\alpha}$ emission.  
As a result of telluric complications, we chose to focus our analysis on the strongest, cleanest feature in each of the spectra, the Br10 line at 1.737 $\mu$m. 
{Figure~\ref{speccomp} shows two 
examples of the Br10 (red) and H${\alpha}$ (black) lines at similar phases for HR 5907 and HR 7355.  Comparing the heights of the red vs. blue 
peaks of the lines, corresponding optical and IR line profiles mimic each other in relative heights.  For HR 5907, phase $\sim0.45$ 
displays a stronger red peak, while phase $\sim0.55$ shows a stronger blue peak.  This is consistent with the maximum H${\alpha}$ 
emission occurring at phase $\sim0.5$, as shown in Fig. 3 of \citet{Grunhut:2012ab}.  The profile for HR 7355 at 0.75 shows two nearly equal peaks, which is expected given that its H$\alpha$ peak occurs
at phase 0.78 \citep{Rivinius:2013aa}.  We also note that Fig.~\ref{spec} suggests an effect whereby the peaks of the emission in Br$\gamma$ trail behind the pattern observed in Br10.  However, this is probably a result of the telluric contamination in Br$\gamma$.  To further reiterate the similarity of the emission peak behavior in the two wavelength domains, Fig.~\ref{vrvar} shows V/R variations for HR 5907, measured from both H$\alpha$ and Br10 (red asterisks) lines.  The agreement between the two diagnostics is striking.  An analogous plot for HR 7355 was not possible given that we have fewer IR data points, but also because the emission in this star is weaker than for HR 5907, and thus more difficult to precisely measure outside of maximum phases.   }

{Figure~\ref{peakvel} demonstrates that the bulk of the emitting plasma, in both cases, is located far beyond the stellar radius (dotted line) and is comparable with the value found for the peak 
in the optical from H$\alpha$ \citep[dashed lines,][]{Grunhut:2012ab,Rivinius:2013aa}.}
{Although there may appear to be a slight difference in the location of the IR material (the red peaks appear systematically at higher velocities), this may just be a consequence of the low resolution, and higher-quality data are necessary to confirm or contradict this.}
As the circumstellar material responsible for the observed emission is in forced co-rotation, the location of the material is directly related to the projected velocity.  For both stars, the bulk of the material is thus found at $\sim$ 2 R$_\star$, coincident with the location of the Kepler corotation radius.  This result agrees with MHD simulations \citep[see, e.g.,][Fig. 9]{Ud-Doula:2008aa}.  

Based on this evidence, we assert the equivalency of the hydrogen recombination lines in the IR with those in 
the optical for diagnosing the presence of a magnetic field.  

Recently, spectropolarimetric observations of one of the IR emission magnetic candidates of \citet{Eikenberry:2014aa} indeed revealed a strong magnetic field \citep{sikora}, quite similar to that of $\sigma$ Ori E.  This result provides confirmation that massive stars with magnetic fields can be detected through IR emission features, although no IR variability study has been performed in this case.  Additional comparison of the IR and optical profiles demonstrates that there is far less contribution from absorption profiles in the IR, revealing more clearly emission from the magnetosphere.  
This attribute is significant, considering that \citet{Najarro:2011aa} have shown that the hydrogen lines in the IR are more sensitive to low mass-loss rates, 
detecting rates ten times lower than measurable by H${\alpha}$ emission, and is particularly important if the physical parameters suggest a detectable magnetosphere 
\citep[given the guidelines determined by ][]{ud-Doula:2002aa,Petit:2013aa}, but optical diagnostics fail.  
For these situations, IR observations may be more effective at detecting and characterizing the properties of circumstellar material, especially for low-density environments.  

{Given the positive detection of IR emission features in known magnetic stars, we have already begun subsequent work to obtain higher resolution data to study the structure of known magnetospheres
with the aim to precisely determine the physical parameters of the emitting plasma, using modern techniques such as tomography \citep{Grunhut:2013ab}.  We have also obtained data to study the effects mentioned in the previous paragraph, whereby lower density environments, while undetected in optical features, may be observed in IR features.  This study would then allow characterizing
a whole new set of magnetospheres with varied physical parameters,
which in turn would motivate further development of magnetic models and theory.   We also aim to understand the IR line profiles of known magnetic O-stars and to determine what information may be obtained from such features so that we can better understand the interaction between the strong stellar wind and the magnetic field.  Fundamentally, the research presented here serves as the initial step towards modernizing the study of magnetic massive stars, with these few example studies constituting the beginning of a larger shift in perspective.  }

%Intuitively, utilizing features in the IR may not seem useful with the decline of flux for hot stars in this wavelength 
%region of the blackbody curve, however, in the presence of even marginal amounts of extinction \citep[$A_V > 5.2$ mag,][]{Hanson:1994aa}, the IR remains largely unaffected.  

As many hot stars are hidden in star-forming regions and other areas that are highly obscured from 
optical and UV observations, such as the Galactic center \citep[see, e.g.,][]{Wachter:2010aa,Gvaramadze:2010aa}, 
IR spectroscopy presents a solution for both identifying and studying new magnetic hot stars and their circumstellar environments.
In fact, studying magnetic stars in the IR may (with higher resolution and S/N) be able to determine the consequences of 
the environment on the incidence of magnetism, that is, various types stellar clusters, the Galactic center vs. within the Galactic disk, high vs. low metallicity, etc.  

In the future, this technique can be used to identify targets for IR spectropolarimetry, both Galactic and extragalactic, to identify and determine physical characteristics of the magnetosphere, and hence infer some properties of the magnetic field itself.  These new technical capabilities will also be crucial in searching for young O-type stars that only become visible once their natal material is shed, providing unprecedented clues to the origin of magnetism in the most massive stars.  Clearly, these types of observations are imperative to reveal and characterize magnetic massive stars, particularly undiscovered objects in heavily shrouded areas.

\section{Conclusions}

While the study presented here contains results from only two stars extracted from low-resolution data, the concepts and findings are applicable 
to the entire field of magnetic hot stars.  In particular, we stress that IR spectra, particularly the H-band Brackett series, are ideal for studying
hot-star magnetospheres, given the lower contribution of stellar flux at these wavelengths.  We can detect this material and its
variability even with low-resolution specctra.  The identified IR spectral features match those found in optical spectra well, including the derived location of the detected material.  {We are currently extending this work to higher resolution spectra with the aim to study known magnetic stars, with the intent to apply this knowledge to candidate stars in regions inaccessible to UV and optical observations. }

IR spectroscopy may be an essential tool for detecting and studying lower density environments around magnetic stars, for detecting magnetic candidates in the Galactic center
and star-forming regions, for studying young OB stars, and for increasing the number of known magnetic stars to determine basic
formation properties.  There is an urgency to understand and develop tools and techniques in the IR not only because massive star magnetism is a rapidly growing field, but also because of the innovative development of IR spectropolarimeters, such as SPIRou \citep{Artigau:2014aa} and CRIRES+ \citep{Dorn:2014aa}, which will require in-depth knowledge of the features and behavior of magnetic diagnostics in the IR and strategies for best exploiting the results to determine physical properties.  Ultimately, observations and models of these magnetospheres in multiple wavebands are crucial for understanding hot stars, their winds, and the effect of magnetic fields on the star and its surroundings.

\begin{acknowledgements}
We thank J. Sundqvist for useful discussions. We also thank Th. Rivinius for supplying UVES spectra of HR 7355. MK acknowledges financial support from GA\,\v{C}R (grant 14-21373S) and from the European Structural Funds grant for the Centre of Excellence "Dark Matter in (Astro)particle Physics and Cosmology".  The Astronomical Institute Ond\v{r}ejov is supported by the project RVO:67985815.  OSIRIS is a collaborative project between the Ohio State University and Cerro Tololo Inter-American Observatory (CTIO) and was developed through NSF grants AST 90-16112 and AST 92-18449. 

%CTIO is part of the National Optical Astronomy Observatory (NOAO), based in La Serena, Chile. NOAO is operated by the Association of Universities for Research in Astronomy (AURA), Inc. under cooperative agreement with the National Science Foundation.
\end{acknowledgements}

\bibliographystyle{aa}
\bibliography{IRbiblio}

\end{document}